\documentclass[12pt,oneside]{revtex4-2}
\usepackage{amssymb}
\usepackage{footmisc}
\usepackage{natbib}
\usepackage{anysize}
\usepackage{graphicx}
\usepackage{amsmath}
\usepackage{wasysym}

\usepackage{setspace}
\usepackage[bookmarks = false]{hyperref}
\usepackage{color}
\usepackage{subfigure}
\usepackage{multirow}
\usepackage[T1]{fontenc}
\renewcommand{\baselinestretch}{1.5}
\renewcommand{\andname}{\ignorespaces}
\marginsize{2.3cm}{2.3cm}{1.5cm}{1.5cm}
\usepackage{color}
\definecolor{SH}{RGB}{0,0,200}
\definecolor{SH2}{RGB}{0,200,200}

\usepackage{setspace}
\makeatletter

\makeatletter

\begin{document}

\title{Telecom-band integrated multimode photonic quantum memory}

\author{Xueying Zhang$^{1,\#}$}
\author{Bin Zhang$^{3,\#}$}
\author{Shihai Wei$^{1}$}
\author{Hao Li$^{4}$}
\author{Jinyu Liao$^{1}$}
\author{Cheng Li$^{1}$}
\author{Guangwei Deng$^{1,5}$}
\author{You Wang$^{1,6}$}
\author{Haizhi Song$^{1,6}$}
\author{Lixing You$^{4}$}
\author{Bo Jing$^{1}$}
\author{Feng Chen$^{3,\dagger}$}
\author{Guang-Can Guo$^{1,5}$}
\author{Qiang Zhou$^{1,2,5,\ast}$}

\affiliation{$^1$Institute of Fundamental and Frontier Sciences, University of Electronic Science and Technology of China, Chengdu 610054, China}
\affiliation{$^2$School of Optoelectronic Science and Engineering, University of Electronic Science and Technology of China, Chengdu 610054, China}
\affiliation{$^3$School of Physics, State Key Laboratory of Crystal Materials, Shandong University, Jinan 250100, China}
\affiliation{$^4$Shanghai Institute of Microsystem and Information Technology, Chinese Academy of Sciences, Shanghai 200050, China}
\affiliation{$^5$CAS Key Laboratory of Quantum Information, University of Science and Technology of China, Hefei 230026, China}
\affiliation{$^6$Southwest Institute of Technical Physics, Chengdu 610041, China}
\affiliation{$^{\#}$ These authors contributed equally to this work.}
\affiliation{Corresponding authors. Email: $^{\dagger}$drfchen@sdu.edu.cn (FC), $^{\ast}$zhouqiang@uestc.edu.cn (QZ)}

\maketitle


\textbf{
	\\
 Telecom-band integrated quantum memory is an elementary building block for developing quantum networks compatible with fiber communication infrastructures. Towards such a network with large capacity, an integrated multimode photonic quantum memory at telecom band has yet been demonstrated. Here we report a fiber-integrated multimode quantum storage of single photon at telecom band on a laser-written chip. The storage device is a fiber-pigtailed Er$^{3+}$:LiNbO$_{3}$ waveguide and allows a storage of up to 330 temporal modes of heralded single photon with 4-GHz-wide bandwidth at 1532 nm and a 167-fold increasing of coincidence detection rate with respect to single mode. Our memory system with all-fiber addressing is performed using telecom-band fiber-integrated and on-chip devices. The results represent an important step for the future quantum networks using integrated photonics devices.}

\vspace{0.5cm}


\renewcommand\section[1]{
	\textbf{#1}
}
\clearpage
\section{\\Introduction}
\\Photonic quantum memories\cite{lvovsky2009optical,simon2010quantum}, allowing reversible mapping of quantum states of light onto matter, are an indispensable component for quantum repeater based long-distance quantum communication\cite{sangouard2011quantum} and distributed quantum networks\cite{kimble2008quantum,wehner2018quantum,simon2017towards,wei2022towards}. Optical waveguides based photonic quantum memory devices connect with other integrated quantum devices, such as quantum light source\cite{uppu2020scalable,harris2014integrated,wang2016chip}, photonic circuit\cite{pitsios2017photonic}, and single photon detector\cite{korzh2020demonstration,marsili2013detecting,you2013jitter,pernice2012high,you2020superconducting}, which will open the way to integrated multifunctional quantum architectures. Rare-earth-ion-doped optical waveguides are excellent candidates for the development of on-chip quantum memory devices, due to compactness, scalability and enhanced light-matter interaction\cite{sinclair2017properties}. Substantial efforts have been devoted to manufacturing on-chip storage devices through various methods, such as Ti$^{4+}$ in-diffusion\cite{dierolf2001spectral,sinclair2010spectroscopic}, focused-ion-beam milling\cite{miyazono2016coupling}, and femtosecond laser micromachining (FLM)\cite{gattass2008femtosecond}. Based on devices from these methods, photonics quantum memories have already demonstrated in different systems\cite{saglamyurek2011broadband,saglamyurek2012conditional,askarani2019storage,liu2022demand,craiciu2019nanophotonic,craiciu2021multifunctional,corrielli2016integrated,sinclair2014spectral,seri2019quantum,su2022demand,zhong2017nanophotonic,zhu2020coherent,liu2020reliable,liu2020demand,seri2018laser,rakonjac2022storage,zhu2022demand,dutta2022atomic}. Recently, erbium-ion-doped waveguides, thanks to the optical translation of $^4$I$_{15/2}\leftrightarrow^4$I$_{13/2}$ at telecom band, have been used to realize integrated quantum memory at 1.5 $\mu$m wavelength\cite{askarani2019storage,liu2022demand,craiciu2019nanophotonic,craiciu2021multifunctional,saglamyurek2016multiplexed,saglamyurek2015quantum,jin2015telecom}. Towards a quantum repeater with large capacity, on-chip quantum memories with broadband\cite{saglamyurek2011broadband,saglamyurek2012conditional,askarani2019storage,sinclair2014spectral} and multimode\cite{craiciu2019nanophotonic,sinclair2014spectral,seri2019quantum,su2022demand} properties have been demonstrated. For practical integrated quantum memory, an effective method is to develop fiber-integrated quantum storage chips\cite{liu2022demand,rakonjac2022storage}, which can directly interconnect to current fiber system, thus facilitating the use of integrated devices in quantum networks. To date, such a fiber-integrated quantum storage chip with multimode capacity at telecom band has yet been demonstrated, especially a broadband storage of non-classical light, which is a key step towards large-scale high-rate quantum networks compatible with existing telecom infrastructure.\\
\indent
Here we demonstrate a telecom-band integrated multimode storage in a Er$^{3+}$:LiNbO$_{3}$ waveguide. The laser-written waveguide fabricated by FLM is directly coupled to a single mode fiber-pigtail via an optical collimator at each end, guaranteeing the compatibility with fiber communication system. An on-chip quantum memory system is demonstrated with the fiber-integrated waveguide based on atomic frequency comb (AFC) protocol. With a 4-GHz-wide AFC, we experimentally realize a multimode quantum storage of 330 temporal modes of heralded single photons at 1532 nm with a 167-fold increasing of coincidence detection rate compared to the single mode. Our quantum memory device paves the way for integrated memory based quantum networks compatible with infrastructures of optical communication.

\section{\\Results}
\\The storage device is a type-III waveguide fabricated in a wafer of Er$^{3+}$:LiNbO$_{3}$ crystal by FLM\cite{zhang2020second}. Figure 1A shows a picture of the bulk crystal of Er$^{3+}$:LiNbO$_{3}$ used in our demonstration. The dopant concentration of Er$^{3+}$ ions is 0.1 mol\%. The bulk crystal is further divided into wafers with a dimension of 10 × 10 × 0.5 mm$^{3}$ along the x × y × z axes (z-cut), in which a laser-written waveguide is fabricated along the y axes. The coupling structure between the laser-written waveguide and single mode fibers is shown in Fig. 1B. The Er$^{3+}$:LiNbO$_{3}$ crystal is glued on a copper heat sink, and two optical collimators with single mode fiber-pigtails are glued $\sim$1.5 mm from the end face of the waveguide. Figure 1C shows the microscope image of the end-face of our waveguide. The fiber-integrated device is placed in a dilution refrigerator with a temperature of 13 mK. A total optical transmission efficiency is 26\% through the entire cryogenic setup. We measure its absorption spectrum, as shown in Fig. 1D. The measured inhomogeneous spectral width of Er$^{3+}$ ions is 180 GHz at $^4$I$_{15/2}\leftrightarrow^4$I$_{13/2}$ transition. The preparation of AFC requires frequency-selective population transfer of Er$^{3+}$ ions from the $^4$I$_{15/2}$ electronic ground level ($\vert g\rangle$) through the $^4$I$_{13/2}$ excited level ($\vert e\rangle$) onto an auxiliary level ($\vert s\rangle$), i.e., long-lived Zeeman sublevel (see the inset of Fig. 1D). We characterize the lifetimes of Zeeman sublevels of Er$^{3+}$ ions by measuring the spectral hole burning - see Supplementary Materials note S2 for details of the setup. Figure 1E shows the decay of central spectral hole depth as a function of time delay, which is fitted by a double exponential polynomial curve $\varDelta \alpha(t)= \varDelta \alpha_ae^{-(t/\tau_a)}+\varDelta \alpha_be^{-(t/\tau_b)}$, where $\varDelta \alpha_a$ and $\varDelta \alpha_b$ are relative amplitudes of two population components present in the hole decay, $\tau_a$ and $\tau_b$ are 1/e population lifetimes of the short-decay and the long-decay, respectively. Our fitting results yield $\tau_a$ = 0.55±0.12 s with a weight of 29\% and $\tau_b$ = 32.75±2.31 s with a weight of 71\%, indicating the presence of two different classes of Er$^{3+}$ ions in the lattice with different nuclear spin relaxation times\cite{thiel2014tm}. Moreover, due to the superhyperfine coupling of Er$^{3+}$ electronic states to neighboring $^{93}$Nb and $^{7}$Li nuclear spins\cite{askarani2020persistent,thiel2010optical}, two classes of side-holes are observed in the hole burning spectrum (see the inset of Fig. 1E). In our experiment, we measure the side-hole detunings under different magnetic fields. The fitted results of the detunings show a field-dependent detuning of 1.077±0.039 kHz/G and 1.487±0.019 kHz/G for side holes arising from the $^{93}$Nb and the $^{7}$Li, respectively. More details on the measurement of side holes are given in Supplementary Materials note S2. With a magnetic field of 1.3 T, i.e. 13000 G, the $^{7}$Li caused side holes with 20 MHz detuning coincide with the transparency regions of the AFC with 5 MHz teeth spacing, while the side holes with 12 MHz detuning caused by the $^{93}$Nb and the AFC is not well matched in our experiment.

Our experimental setup of multimode storage is sketched in Fig. 2A. It consists of the generation of heralded single photon, the preparation of AFC quantum memory and the measurement system. Correlated photon pairs are generated by cascaded second-harmonic generation (SHG) and spontaneous parametric down conversion (SPDC) processes in a fiber-pigtailed periodically poled LiNbO$_{3}$ (PPLN) waveguide module pumped by a series of light pulses at 1540.60 nm\cite{zhang2021high,wei2022storage}. For the single mode storage, the PPLN module is pumped by a single laser pulse with a duration of 300 ps within a period of 1 $\mu$s. And for the multimode storage, the PPLN module is pumped by 330 pulses with a separation of 600 ps within a period of 1 $\mu$s. A pair of photons at 1549.25 nm and 1532.05 nm are named idler and signal photons, respectively, and the bandwidth of photon pairs is 60 nm. The idler photons and signal photons are selected with bandwidths of 6.2 GHz and 5.2 GHz by using two dense wavelength division multiplexers (DWDMs) followed by two fiber-Bragg gratings (FBGs). The photons are detected by superconducting nanowire single photon detectors (SNSPDs). A successful detection output of heralding idler photons heralds the presence of heralded signal photons. The two-fold coincidences detection rates of our correlated photon pair source are 344.78±0.59 Hz and 49904.49±7.06 Hz for the single mode and the multimode storage, respectively. For the storage of the heralded single photon, the signal photons are sent to our fiber-integrated Er$^{3+}$:LiNbO$_{3}$ waveguide, in which a 4-GHz-wide AFC is prepared by using a continuous-wave (CW) laser at 1532.05 nm modulated by a phase modulator (PM) with frequency chirped technology\cite{saglamyurek2011broadband}. The signal photons are mapped onto the AFC leading to a collective atomic excitation state, which can be described by a collective Dicke state\cite{de2008solid}
\begin{equation}
|\Psi\rangle=\frac{1}{\sqrt{N}}\sum_{j}^Nc_{j}e^{i2\pi\delta_{j}t}e^{-ikz_{j}}|g_{1}g_{2}...e_{j}...g_{N}\rangle
\end{equation}
where $N$ is the total number of ions, $|g_{j}\rangle$ and $|e_{j}\rangle$ represent the ground and excited states of ion $j$, respectively. $z_{j}$ is the position of ion $j$; $k$ is the wavenumber of the light field; $\delta_{j}$ is the detuning of the atom with respect to the laser frequency; and the amplitudes $c_{j}$ depend on the frequency and on the spatial position of atom $j$. The atomic collective state will rapidly dephase after absorption of photons. Owing to this periodic structure of the absorption profile, the atomic excitations will rephase and the signal photons will be recalled in the same mode after a predetermined time of $\frac{1}{\Delta}$ (where $\Delta$ is the teeth spacing of the prepared AFC). The recalled signal photons are detected by a SNSPD through an optical switch (OS), which acts as a temporal gate to avoid the pump light for AFC preparation damage the SNSPDs and ensure that recalled photons are not contaminated by noise photons stemming from the spontaneous decay of atoms with the excited state. Detection signals of the SNSPDs are analyzed by using a time-to-digital converter (TDC).

We start to tailor a 4-GHz-wide inhomogeneous broadened profile of Er$^{3+}$ ions into a periodical absorptive structure, i.e., AFC. Figure 3A shows a 40-MHz-wide section. The teeth spacing is 5 MHz, corresponding to a storage time of 200 ns. Consequently, we demonstrate the storage of non-classical light with a large time-bandwidth product up to 800 (4 GHz × 200 ns). The signal photons with the full width at half maximum (FWHM) of 300 ps are sent to the AFC memory. Figure 3B shows the transmitted photons and the recalled photons with the predetermined storage time of 200 ns. A system efficiency is calculated to 0.59±0.01\% by the ratio of recalled heralded signal photon counts to all the input ones. Considering the transmission efficiency of the memory device and the spectral filtering of the input photons caused by the slightly smaller bandwidth of the AFC memory, an internal storage efficiency is calculated to 2.83±0.03\%. More details about internal storage efficiency can found in Supplementary Materials note S5 and S6. The measured two-fold coincidences detection rates are 344.78±0.59 Hz and 1.95±0.04 Hz before and after storage, respectively. To assess the non-classical property of the AFC quantum memory, we measure the second-order cross-correlation function $g_{s,i}^{(2)}(t)$
\begin{equation}
g_{s,i}^{(2)}(t)=\frac{P_{si}(t)}{P_{s}(t)\cdot P_{i}}
\end{equation}
where $t$ is the storage time, $P_{si}(t)$ is the probability of the two-fold coincidence detection between idler photons and signal photons triggered by the system clock, and $P_{s}(t)$($P_{i}$) is the probability of detecting the signal (idler) photons. The $g_{s,i}^{(2)}(0)$ for correlated photon pair source is 23.41±0.04 before quantum storage - the red horizontal line in Fig. 3C, far exceeding the classical limit of 2\cite{fasel2004high,tapster1998photon} - the green horizontal dashed line in Fig. 3C. We measure the values of $g_{s,i}^{(2)}(t)$ between idler and recalled signal photons at different storage times ($t$) from 100 ns to 240 ns, as shown in Fig. 3C. It shows that all values of $g_{s,i}^{(2)}(t)$ are around 23 even for the maximal storage time, which demonstrates that the AFC quantum memory preserves the non-classical properties of correlated photon pairs. The slight changes of $g_{s,i}^{(2)}(t)$ are owing to the high signal-to-noise ratio of our AFC quantum memory, which is realized by synchronizing the storage process with the system clock\cite{wei2022storage}. Furthermore, with the increased source-pump power and repetition rate, we measure the heralded second-order autocorrelation function $g_{i:s,s}^{(2)}(t)$. The measured values are 0.23±0.01 and 0.22±0.06 before and after storage, respectively. We further measure the unheralded second-order autocorrelation function $g_{s,s}^{(2)}(t)$ of the signal photons. The measured values are 1.64±0.01 and 1.58±0.35 before and after storage, respectively. These results further indicate that the AFC quantum memory keeps the single photon purity and the spectral purity of photons.

We then establish a multiple temporal mode operation of on-chip quantum memory with the storage time of 200 ns. As shown in Fig. 4A, there are 330 distinguishable temporal modes of signal photons transmitted and recalled from the AFC quantum memory. The FWHM width of each temporal mode is 300 ps with a separation of 600 ps. The measured two-fold coincidences detection rates are 49904.49±7.06 and 326.70±0.57 Hz before and after storage, respectively. Note that the dropping for the temporal modes with higher index number is due to the 50-ns dead time of SNSPDs. To assess the non-classical properties of all stored modes throughout the AFC storage process, we obtain two sets of 330 × 330 values of $g_{s,i}^{(2)}(0)$ for photon pairs source and $g_{s,i}^{(2)}(t=200\ $ns) between the idler and recalled signal photons, respectively. Figures 4B and 4C present parts of the second-order cross-correlation function corresponding to different temporal modes before and after quantum storage (the full 330 × 330 arrays of the second-order cross-correlation function are shown in Supplementary Materials note S7). We find that, for the 330 pairs of correlated temporal modes, the average $g_{s,i}^{(2)}(0)$ is 19.39±0.01 and the average $g_{s,i}^{(2)}(t=200\ $ns) is 18.93±0.04, which demonstrates that the non-classical properties of all temporal modes still maintain after the AFC quantum storage. The crosstalk between different temporal modes is assessed via measuring the second-order cross-correlation function between uncorrelated temporal modes\cite{pu2017experimental}. As expected, for all uncorrelated temporal modes, the average second-order cross-correlation function is 0.99±0.01 and the one after quantum storage is 1.00±0.01, which verifies the crosstalk is negligible.  
\\
\section{\\Discussion}
\\In the present work, we have demonstrated an fiber-integrated multimode quantum memory based on a laser-written Er$^{3+}$:LiNbO$_{3}$ waveguide. A time-bandwidth product of 800 is achieved with a storage bandwidth of 4 GHz and a storage time of 200 ns. We have shown that the non-classical correlations between idler photons and recalled signal photons at telecom band are well maintained after the retrieval. We have shown a quantum storage of 330 temporal modes of single photon at 1532 nm with a 167-fold increasing in coincidence detection rate compared to the single mode, and measured the crosstalk between uncorrelated temporal modes which is negligible. Our broadband multimode quantum storage will open the way to high-rate quantum network compatible with fiber communication infrastructures.

Despite these important results, several upgrades are needed towards a functional device for quantum network. The storage time of our demonstration is limited by the setups for the AFC preparation. For instance, the linewidth of our AFC pump laser is around 100 kHz with a frequency drift of MHz level, which prepares the AFC with a minimal teeth spacing of several MHz, resulting in a maximal storage time of hundreds of nanoseconds. Therefore, it needs to update the laser system by using an ultra-stable laser system to prepare the AFC\cite{legero20171,snigirev2023ultrafast}. From the material side, it is necessary to extend the optical coherence time of the Er$^{3+}$ ions in the waveguide (more details on optical coherence time see Supplementary Materials S3), which could be achieved by optimizing the dopant concentration of the Er$^{3+}$ ions\cite{thiel2010optical}. On the other hand, the observed two classes of side holes influence the performance of the AFC memory which are from the superhyperfine coupling of Er$^{3+}$ electronic states to neighboring nuclear spins of the $^{93}$Nb and the $^{7}$Li. Ideally the influence of these side holes could be eliminated by setting the transparency regions of the prepared AFC to coincide with the side holes with applying an optimized magnetic field\cite{etesse2021optical}. In our experiment, the storage time of 200 ns corresponds to an AFC with a teeth spacing of 5 MHz. With a magnetic field of 1.3 T, the transparency region of such AFC coincides with the side holes with the 20 MHz detuning from the neighboring $^{7}$Li. Hereby the $^{93}$Nb caused side holes may inescapably influence the performance of our memory. Possible method to avoid the side-hole effect is to implement quantum memory with other protocols\cite{ma2021elimination}, or to realize the AFC with specified storage times, such as 250 ns and 500 ns. Furthermore thanks to the on-chip integration property of our demonstration it is also feasible to further improve the performance of our Er$^{3+}$:LiNbO$_{3}$ memory, for instance integrated with an on-chip microcavity for impedance-matched quantum memory\cite{afzelius2010impedance,sabooni2013efficient} and with electrodes for the possible on-demand recall\cite{craiciu2021multifunctional,zhu2022demand,hastings2006controlled}. It is also worth further exploring a possible $\Lambda$ -like hyperfine energy-level system in our device to perform an on-demand recall based on the spin wave storage\cite{gundougan2015solid,jobez2015coherent}, which could be a particular challenge in Er$^{3+}$:LiNbO$_{3}$ device since the Er$^{3+}$ ions are the Kramers ions\cite{ranvcic2018coherence}. Finally, towards memories with larger capacities, a promising avenue is to combine multiple degrees of freedom. In addition to multiple temporal mode operation\cite{afzelius2009multimode}, more spectral channels could be prepared in the large inhomogeneous broadening of the Er$^{3+}$:LiNbO$_{3}$ waveguide with optical frequency combs\cite{wei2022storage}. Moreover, more spatial channels could also be exploited in an on-chip quantum memory by fabricating several waveguides in the crystal.

In conclusion, our approach combines the reliability of a fiber-integrated device compatible with the fiber telecom infrastructure, the broadband multiplexed storage properties, and promising laser-written components. Based on our demonstration of 247 multimode storage for heralded single photons at telecom wavelength - a 167-fold 248 increasing compared to the single mode, next steps could be the storage of qubit, the creation 249 of entanglement between remote memory chips, the implementation of feed-forward control 250 and the on-demand recall. The combination of photon pair sources and integrated memories with higher multimode capacity should allow the realization of a high-rate quantum repeater protocol towards a large-scale quantum network\cite{simon2007quantum}. Our system presents a fundamental building block towards the realization of a quantum repeater with large capacity, scalability and compatibility of fiber communication infrastructure.

\section{\\Materials and Methods}\\
\textbf{Waveguide fabrication and fiber pigtailing.} Our integrated storage device is a type-III waveguide fabricated in a wafer of Er$^{3+}$:LiNbO$_{3}$ crystal (along crystal y-axis) utilizing FLM technology. The wafer of crystal - with a dimension of 10 × 10 × 0.5 mm$^3$ along the x × y × z axes (z-cut) - is cut from a bulk 0.1 mol\% Er$^{3+}$:LiNbO$_{3}$ crystal, which is grown by Shanghai Institute of Optics and Fine Mechanics, Chinese Academy of Sciences. The waveguide is fabricated by a fiber femtosecond laser (FemtoYLTM-25, YSL Photonics). The central wavelength, pulse duration, and repetition rate are 1031 nm, 400 fs, and 25 kHz, respectively. A microscopic objective (50×, N.A. = 0.67, Sigma) has been utilized to focus femtosecond laser into the wafer of Er$^{3+}$:LiNbO$_{3}$ crystal along z-axis (c-axis). The focusing depth is around 180 µm, and the pulse energy of femtosecond laser on sample surface is around 0.26 $\mu$J. The Er$^{3+}$:LiNbO$_{3}$ sample has been placed onto a motor-driven translation stage, and the scanning velocity along crystal y-axis is set as 3 mm/s. The polarization of femtosecond laser is perpendicular to the crystal y-axis. As a result, a depressed-cladding waveguide is fabricated as shown in Fig. 1C. After femtosecond laser micromachining, thermal annealing (annealed at 200 ℃ for 0.5 h in air) has been applied to improve the guiding properties of waveguide. To realize on-chip quantum memory compatible with fiber communication infrastructure and to avoid challenges of free-space coupling, the waveguide is packaged into a fiber-pigtailed device. Two optical collimators (FCSS-G-155-1.8-A-1.1-025, Femto Technology, Xi’an) with single mode fiber-pigtails are coupling with the waveguide. The collimator is in cylinder shape with a length of 9 mm, a diameter of 1.4 mm, and a beam waist of around 50 $\mu$m. The operating wavelength range of optical collimators is 1510-1590 nm. Before gluing the collimators, they are held on a motorized stage with a moving resolution of 500 nm (A10-60, HUAVE). The distance between the collimators and the waveguide is monitored by using a microscope. The transmission efficiency of the coupling process is measured by using a laser light at 1570 nm - out of the absorption spectrum of the Er$^{3+}$:LiNbO$_{3}$ waveguide. The two optical collimators are glued on the heat sink when the measured transmission reaches maximum, i.e., a transmission efficiency of 40\% in our demonstration. The loss is caused by the coupling mode mismatch, the waveguide propagation loss, imperfect waveguide facets and so on. After mounting the fiber-pigtailed device into the dilution refrigerator and the cooling the temperature down to 13 mK, the total transmission efficiency through the entire cryogenic setup drops to 26\% due to the fiber splicing and coupling misalignment in the cooling process. In perspective, we believe that one good strategy to improve the device transmission efficiency is to optimize the waveguide irradiation recipe to obtain a mode field distribution that better matches the collimator. 
\\
\textbf{AFC quantum memory.} The fiber-pigtailed laser-written Er$^{3+}$:LiNbO$_{3}$ waveguide is cooled down to 13 mK in a dilution refrigerator (LD400, Bluefors) integrated with superconducting magnetic field of 1.3 T parallel to the c axis of crystals. We use a frequency-doubled laser diode (TOPTICA, CTL 1550) to generate the CW laser at 1532.05 nm for the AFC preparation. The light is modulated to a continuous frequency chirped laser through frequency modulation, which is implemented using a lithium niobate PM driven by a 25 GS/s an arbitrary waveform generator (AWG) combining with sideband-chirping technique. The frequency chirping extends from -2 GHz to 2 GHz, thus efficiently preparing a 4-GHz-wide AFC memory. The pump power that we inject in the waveguide during the AFC creation is controlled to 28 $\mu$W by a variable optical attenuator (VOA), and the pump time is set to 200 ms by an OS. During this optical pump duration time, the frequency chirp cycle is repeated 1250 times. To ensure the recalled photons are not contaminated by noise photons stemming from the spontaneous decay of atoms with the excited state, a wait time of 20 ms is applied after the optical pumping by controlling the state of the OS before the dilution refrigerator. Meanwhile all $g_{s,i}^{(2)}(t)$ results are recorded by using a two-fold coincidence detection triggered by the system clock. Following the wait time, it is 280 ms for storage and recall of the signal photons in the 500 ms measurement cycle. Before the signal photons enters the waveguide, the OS before the measurement system is turned off to protect the SNSPD.
\\
\textbf{Heralded single photon source.} Correlated photon pairs are obtained by pumping a fiber-pigtailed PPLN waveguide module with a train of laser pulses through cascaded second-harmonic generation and spontaneous parametric down conversion processes. The laser pulses are prepared by using a lithium niobite intensity modulator (IM) to externally modulate a CW laser at 1540.6 nm, which is generated from a narrow-linewidth semiconductor laser (PPCL300, PURE Photonics). The external modulation signals of the IM are derived from the pulsed electrical signals generated by an AWG and amplified to around   of the IM by a microwave amplifier. To ensure the optimal extinction ratio of light pulse, we use a 99:1 beam splitter (BS) combined with a photodetector to obtain a feedback electrical signal to the IM. These laser pulses are amplified by an erbium-doped fiber amplifier (EDFA) and then filtered out the fluorescence noise by a DWDM with a central wavelength at 1540.60 nm and a passband width of $\sim$100 GHz. The power of laser pulses is attenuated, and monitored by a VOA and a 90:10 BS with an optical power meter (OPM), respectively. To maximize the efficiency of phase matching before the PPLN waveguide module, a polarization beam splitter (PBS) and a polarization controller (PC) are used to manipulate the polarization of pump laser. The bandwidth of correlated photon pairs is $\sim$60 nm generated by the PPLN module. After the module, an isolator is used to reject the residual second-harmonic photons at 770 nm. Finally, heralded signal photons at 1532.05 nm (heralding idler photons at 1549.25 nm) are spectrally filtered by a DWDM with a passband of $\sim$100 GHz and a FBG with a passband of $\sim$5.2 ($\sim$6.2) GHz. The idler photons are directly detected by a SNSPD (P-CS-6, PHOTEC Corp., detection efficiency of 59\% and 70 Hz of dark counts). Meanwhile, signal photons are sent to the fiber-integrated waveguide for storage and reemission, and detected by another SNSPD (P-CS-6, PHOTEC Corp., detection efficiency of 76\% and 70 Hz of dark counts). Detection signals are finally analyzed by a TDC (quTAG, qutools), which records the detection counts of the idler/signal channel and coincidence counts between the two channels triggered by the system clock (see Supplementary Information note S4 for more details of heralded single photon source). 
\\
\textbf{Calculation of $g_{s,i}^{(2)}(t)$, $g_{i:s,s}^{(2)}(t)$, and $g_{s,s}^{(2)}(t)$.} The correlation between the idler and the signal photons is investigated by measuring the second-order cross-correlation function $g_{s,i}^{(2)}(t)$,
\begin{equation}
g_{s,i}^{(2)}(t)=\frac{C_{si}(t)\times N}{C_{s}(t)\times C_{i}}
\end{equation}
where $N$ is the experimental trials for counting effective events with the integrated measurement time window; $C_{si}(t)$ is the two-fold coincidence counts between idler and signal photons triggered by the system clock; $C_{s}(t)$ ($C_{i}$) is the detection counts of the signal (idler) photons.

The signal photons are split by a 50:50 fiber beam splitter (FBS) and labled as $s_1$ and $s_2$ for different outputs of FBS. The single photon purity is evaluated by measuring the heralded second-order auto-correlation function $g_{i:s,s}^{(2)}(t)$,
\begin{equation}
g_{i:s,s}^{(2)}(t)=\frac{C_{i,s_1,s_2}(t)\times C_i}{C_{i,s_1}(t)\times C_{i,s_2}(t)}
\end{equation}
where $C_{i,s_1,s_2}(t)$ is the three-fold coincidence counts among idler photons, $s_1$, $s_2$, and $C_{i,s_1}(t)$($C_{i,s_2}(t)$) is the two-fold coincidence counts between idler photons and $s_1$($s_2$). All the coincidence measurements are triggered by the system clock.

The spectral purity is characterized by measuring the unheralded second-order auto-correlation function $g_{s,s}^{(2)}(t)$ of the signal photons,
\begin{equation}
g_{s,s}^{(2)}(t)=\frac{C_{s_1,s_2}(t)}{A_{s_1,s_2}(t)}
\end{equation}
where $C_{s_1,s_2}(t)$ is the two-fold coincidence counts between $s_1$ and $s_2$, and $A_{s_1,s_2}(t)$ is the two-fold accidental coincidence counts of signal photons $s_1$ and $s_2$. All the coincidence measurements are triggered by the system clock.

For all the coincidence measurements in this paper, the width of coincidence window is 600 ps. The integrated measurement time is 1000 s and the time-bin width is 10 ps for the measurement of the $g_{s,i}^{(2)}(t)$. The integrated measurement time is 1000 s for the measurement of the $g_{i:s,s}^{(2)}(0)$ and $g_{s,s}^{(2)}(0)$, and the integrated measurement time is > 20 h for the measurement of $g_{i:s,s}^{(2)}(t=200\ $ns) and $g_{s,s}^{(2)}(t=200\ $ns). The time-bin width is 100 ps for the measurement of $g_{i:s,s}^{(2)}(t)$ and $g_{s,s}^{(2)}(t)$.

\section{\\Data availability}\\
All data needed to evaluate the conclusions in the paper are present in the paper and/or the Supplementary Materials.

\section{\\Acknowledgments}\\
This work was supported by the National Key Research and Development Program of China (Nos. 2018YFA0307400, 2018YFA0306102), National Natural Science Foundation of China (Nos. 61775025, 91836102, U19A2076, 12004068, 12174222), Innovation Program for Quantum Science and Technology (No. 2021ZD0301702), Sichuan Science and Technology Program (Nos. 2021YFSY0062, 2021YFSY0063, 2021YFSY0064, 2021YFSY0065, 2021YFSY0066), China Postdoctoral Science Foundation (Nos. 2020M683275, 2021T140093), Natural Science Foundation of Shandong Province (No. ZR2021ZD02).

\section{\\Author contributions}\\
 FC, GG and QZ and conceived and supervised the project. XZ, SW, JL, and CL contributed to the mounting and adapting of the sample to the cryostat. XZ and BZ mainly carried out the experiment and collected the experimental data with help of other authors. XZ, BJ and QZ analyzed the data. HL and LY developed and maintained the SNSPDs used in the experiment. HS and YW provide the temperature control system for the PPLN module. XZ and QZ wrote the manuscript with inputs from all other authors. All authors have given approval for the final version of the manuscript.

\section{\\Competing interests}\\
\\The authors declare that they have no competing interests.

\bibliography{myref}

\clearpage
\begin{figure}[ht]
\centering
\includegraphics[width=16cm]{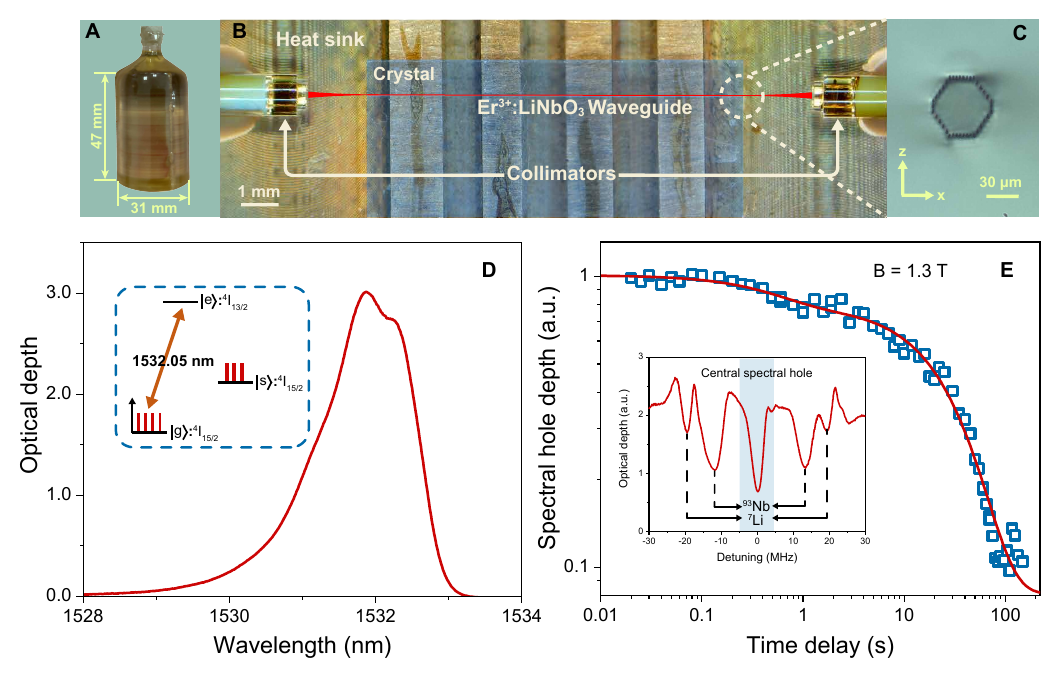}
\caption{\textbf{Preparation and calibration of Er$^{3+}$:LiNbO$_{3}$ waveguide.} (A) Picture of the bulk crystal of Er$^{3+}$:LiNbO$_{3}$. (B) Picture of fiber-integrated laser-written Er$^{3+}$:LiNbO$_{3}$ waveguide. The red line is for guiding eyes. Two optical collimators with single mode fiber-pigtails are used to couple with the waveguide. (C) Picture of the cross section of the waveguide. (D) Absorption spectrum of Er$^{3+}$:LiNbO$_{3}$ waveguide with level scheme - the inset. (E) Spectral hole depth decays with the waiting time with a magnetic field of 1.3 T along c-axis. The experimental data is fitted by double exponential polynomial, from which we extract two lifetimes   and   of 0.55±0.12 s and 32.75±2.31 s, respectively. The weights of   and   decays are 29\% and 71\%, respectively. The inset gives a 60-MHz-wide section of a spectral hole.}
\label{fig:1}
\end{figure}

\clearpage
\begin{figure}[ht]
\centering
\includegraphics[width=16cm]{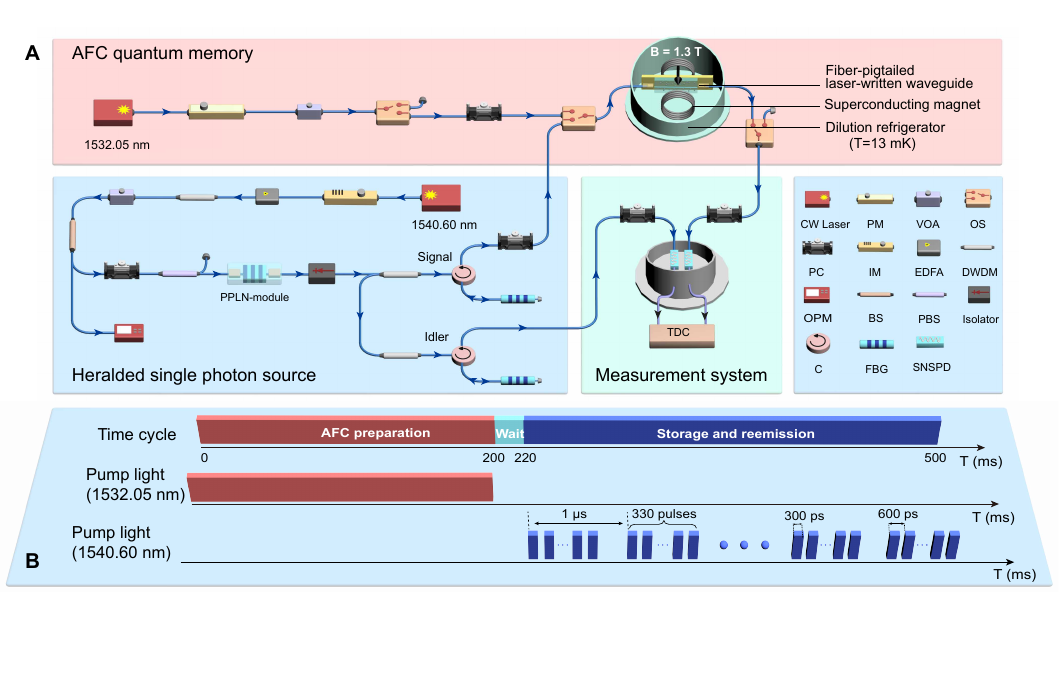}
\caption{\textbf{Experimental setup.} (A) Schematics of the setup. Heralded single photon is obtained based on cascaded second-harmonic generation and spontaneous parametric down conversion processes in a fiber-pigtailed periodically poled LiNbO$_{3}$ (PPLN) waveguide module. Signal photons are directly transmitted into a fiber-integrated laser-written Er$^{3+}$:LiNbO$_{3}$ waveguide placed in a dilution refrigerator. The signal photons are recalled with a preprogrammed storage time. Photons from signal and idler channels are detected by two superconducting nanowire single photon detectors (SNSPDs). A time-to-digital converter (TDC) is used to analyze the detection signals by recording the single and coincidence counts of photons from both channels. CW laser: continuous-wave laser; PM: phase modulator; VOA: variable optical attenuator; OS: optical switch; PC: polarization controller; IM: intensity modulator; EDFA: erbium-doped fiber amplifier; DWDM: dense wavelength division multiplexer; OPM: optical power meter; BS: beam splitter; PBS: polarization beam splitter; C: circulator; FBG: fiber Bragg grating. (B) Experimental timing sequence. The sequence is continuously repeated during the experiment. A whole time period is 500 ms - 200 ms for preparing AFC, 20 ms for waiting, 280 ms for storing signal photons.}
\label{fig:2}
\end{figure}

\clearpage
\begin{figure}[ht]
\centering
\includegraphics[width=16.4cm]{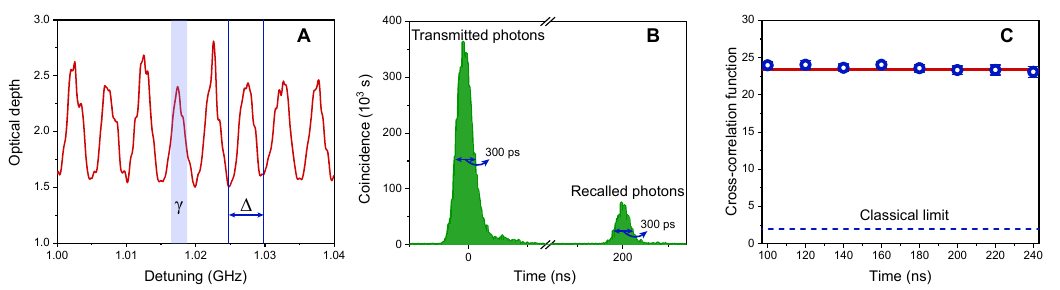}
\caption{\textbf{Characterization of the quantum memory.} (A) A 40-MHz-wide section of the 4-GHz-wide AFC. The teeth spacing $\Delta$ is 5 MHz, corresponding to storage time of 200 ns. The linewidth of the teeth $\gamma$  is ~2.5 MHz. (B) Temporal coincidence histogram between idler photons and signal photons after quantum storage - triggered by the system clock. The predetermined storage time is 200 ns, and the full width at half maximum (FWHM) of a temporal mode is 300 ps. (C) Second-order cross-correlation function $g_{s,i}^{(2)}(t)$ between idler photons and signal photons with different predetermined storage times. The red horizontal line shows the value of 23.41±0.04 of the $g_{s,i}^{(2)}(0)$ for the photon pair source alone before quantum storage. The $g_{s,i}^{(2)}(t)$ after the storage of signal photons still remains around 23. The detection windows of idler and signal channel are set to 600 ps. The classical threshold is reported as a bule horizontal dashed line. The integration time is 1000 s and the time-bin width is 10 ps. The error bars are evaluated from the counts assuming Poissonian statistics.}
\label{fig:3}
\end{figure}

\clearpage
\begin{figure}[ht]
\centering
\includegraphics[width=16.3cm]{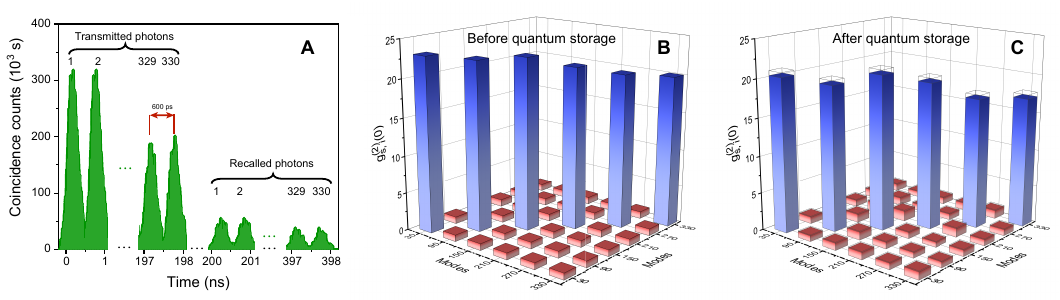}
\caption{\textbf{Results of the quantum storage of 330 temporal modes of heralded single photon.}(A) Temporal coincidence histogram between idler photons and signal photons after quantum storage - triggered by the system clock. The FWHM width of each temporal mode is 300 ps and a separation is 600 ps between adjacent modes. Storage time is set to 200 ns. (B), (C) Second-order cross-correlation function between the idler and signal photons before and after quantum storage. The average second-order cross-correlation function for the 330 pairs of correlated temporal modes before quantum storage (after quantum storage) is 18.93±0.04 (19.39±0.01). The average second-order cross-correlation function of the uncorrelated temporal modes before quantum storage (after quantum storage) is 0.99±0.01 (1.00±0.01). The average value is acquired through Monte Carlo simulation. The detection windows of idler and signal channel of each temporal mode are set to 600 ps. The integration time is 1000 s and the time-bin width is 10 ps. The error bars are evaluated from the counts assuming Poissonian statistics.}
\label{fig:4}
\end{figure}

\clearpage
\section{\centerline{Supplementary Materials}}
\renewcommand{\baselinestretch}{1.5}
\renewcommand\figurename{Fig.S.}
\renewcommand\tablename{TABLE S.}
\renewcommand{\andname}{\ignorespaces}
\setcounter{figure}{0}
\setcounter{table}{0}
\subsection*{Note S1: Absorption spectrum of Er$^{3+}$:LiNbO$_{3}$}
\noindent
Figure S. 1 shows the experimental setup of measuring the absorption spectrum of Er$^{3+}$:LiNbO$_{3}$ waveguide. A broad spectrum light with the bandwidth of 100 nm (the wavelength ranges from 1470 nm to 1570 nm) - from an erbium-doped fiber amplifier (EDFA) - is used to pump the Er$^{3+}$:LiNbO$_{3}$ waveguide. The optical power and the polarization of the input light is controlled by a variable optical attenuator (VOA) and a polarization controller (PC). We set the pump power to 10 $\mu$W, and the spectrum is measured by a spectrum analyzer with a resolution of 0.02 nm. The optical depth corresponding to each wavelength is obtained by the formula: $P_{out}=P_{in}e^{-\alpha L}$\cite{dajczgewand2015optical}, where $P_{out}$ and $P_{in}$ are the optical power after and before the waveguide, $\alpha$ is the absorption coefficient, $L$ is the length of the waveguide, and $\alpha L$ represents the optical depth. The absorption spectrum is shown in Figure S. 1C of main text and the full width at half maximum (FHWM) of inhomogeneous spectral width is around 180 GHz.

\subsection*{Note S2: Zeeman-sublevel lifetime of Er$^{3+}$:LiNbO$_{3}$}
\noindent
We perform spectral hole burning (SHB) to investigate the lifetime of Zeeman sublevels of Er$^{3+}$:LiNbO$_{3}$ at a temperature of 13 mK. The experimental setup is shown in Fig. S2A. A continuous-wave (CW) laser with a wavelength of 1532.05 nm is split into two beams by a 50:50 beam splitter (BS). One is utilized as excitation light to prepare a persistent spectral hole, and another one is utilized as probe light to detect the hole. We shift the frequency of the excitation light by a 1.5 GHz via the phase modulator (PM1) with serrodyne modulation\cite{johnson1988serrodyne}. The excitation time is set to 200 ms by the optical switch (OS1). The OS2 is used to control the time-delay between excitation light and the probe light entering the waveguide. After a varying time-delay, the spectral hole is detected by linearly varying the frequency of the laser light. Frequency sweeps are achieved by the PM2 with serrodyne modulation. The time of frequency sweeps is 10 $\mu$s controlled by an acousto-optic modulator (AOM), and the range is over a 100-MHz-wide frequency window centered at the excitation frequency. During the time for hole preparation, the OS3 are used to avoid the pump laser entering the photodetector (PD). Figure S2B shows the optical depth of Er$^{3+}$ ions over a 60-MHz-wide spectral range centered at 1532.05 nm after applying a magnetic field of 1.3 T. The observed side holes around the central hole result from strong coupling of the Er$^{3+}$ electronic states to neighboring $^{93}$Nb and $^7$Li nuclear spins\cite{askarani2020persistent,thiel2010optical}. The central spectral hole broadened to a FWHM of 4 MHz. The side holes broadened to a FWHM of 6 MHz as the $^{93}$Nb relaxed and then to 3 MHz as the $^7$Li relaxed. The detuning of side holes arising from $^{93}$Nb and $^7$Li are $\sim$12 MHz and $\sim$20 MHz, respectively. To characterize the level splitting influenced by magnetic field, we plot the field-dependent detuning of side-holes at different magnetic field with B//c (see Fig. S2C). Fits yield values of $^{93}$Nb of 1.077±0.039 kHz/G and $^7$Li of 1.487±0.019 kHz/G.

\subsection*{Note S3: Optical coherence time of Er$^{3+}$:LiNbO$_{3}$}
\noindent
The upper limit of the storage time for quantum memory could be the optical coherence time of the Er$^{3+}$:LiNbO$_{3}$, which can be obtained by measuring two-pulse photon echoes. The experimental setup for measuring the two-pulse photon echoes is shown in Fig. S3A. In the experiment, a CW laser with a wavelength of 1532.05 nm is modulated into two-pulse sequence by using an AOM, i.e., a $\pi /2$ pulse and a $\pi$ pulse. The peak power of the pulses is about 8 mW. Figure S3B shows the photon echo area as a function of the time delay $t_{12}$ between the $\pi /2$ pulse and the $\pi$ pulse, which is fitted by $A(t_{12})=A_0e^{-2{(2t_{12}/T_2)}^x}$ , where $A_0$ is the echo area at $t_{12}=0$; $T_2$ is the fitted optical coherence time; $x$ describes the decay shape\cite{mims1968phase}. With the measured experimental results, a fitted $T_2$ of 90.34±1.13 $\mu$s is obtained with a fitted $x$ of 1.93.

\subsection*{Note S4: Properties of heralded single photon source}
\noindent
Figure S4A shows the single detection rates from signal and idler side, which are well fitted with a quadratic polynomial curve\cite{zhang2021high}. Different single detection rate between the signal and idler channel is originated from the different measured efficiency and bandwidth of the filters. The measured efficiency of idler (signal) channel is 17\% (13\%), including the coupling efficiency of 77\% (77\%) for the fiber coupling into a fiber-pigtailed periodically poled LiNbO$_3$ (PPLN) waveguide module, the transmission efficiency of 39\% (25\%) from the output of PPLN to the input of SNSPD, and the detection efficiency of 62\% (76\%) for SNSPD. The bandwidth of the FBG with the central wavelength of idler photons (signal photons) is $\sim$ 6.2 GHz (5.2 GHz). Figure S4B shows the two-fold coincidence detection rate between the idler photons, signal photons triggered by the system clock corresponding to different pump power. The two-fold coincidence detection rate triggered by the system clock is available to eliminate the effects of accidental coincidence counts, which avoids imperfect extinction ratio of light pulse from increasing the coincidence detection rate. Figure S4C shows the second-order cross-correlation function ($g_{s,i}^{(2)}(0)$) as a function of the detected probability of idler photons (P$\rm_{idler}$), which are well fitted with an inversely proportional curve. In the main text, we set the pump power to 36 $\mu$W. For single mode storage, the measured single detection rates of idler and signal channels are 3652.22±1.91 Hz and 2267.77±1.52 Hz, respectively. For multimode storage, the measured single detection rates of idler and signal channels are 466643.66±68.31 Hz and 241836.93±15.55 Hz, respectively. The two-fold coincidences detection rates are 344.78±0.59 Hz and 49904.49±7.06 Hz for the single mode storage and multimode storage, respectively. For single mode storage, the detected probability of idler (signal) photons, i.e., P$\rm_{idler}$ (P$\rm_{signal}$), is measured to 0.63\% (0.41\%) with a $g_{s,i}^{(2)}(0)$ of 23.41±0.04. Combining a measured efficiency of 17\% and the P$\rm_{idler}$, an intrinsic generation probability of idler photons is calculated to 3.71\% in the waveguide. If signal-to-noise-ratio (SNR) of correlated photon pairs is large enough, the theoretical value of $g_{s,i}^{(2)}(0)$ should be 27.98\cite{wei2022storage}, while the experimentally measured value of $g_{s,i}^{(2)}(0)$ is 23.41. The main factor is the existence of spontaneous Raman scattering noise photons generated from the PPLN waveguide module\cite{zhang2021high}. The main parameters of the PPLN waveguide module are shown in Table S1.
\subsection*{Note S5: The internal storage efficiency of AFC}
\noindent
To analyze the internal efficiency of our quantum memory based on the fiber-pigtailed laser-written Er$^{3+}$:LiNbO$_{3}$ waveguide, we define a system efficiency as $\eta_s=N_{out}/N_{in}$ , where $N_{out}$ and $N_{in}$ is the detection rate of the recalled signal photons and input signal photons on condition with heralding idler photons. The system efficiency contains the transmission efficiency ($\eta_t$), the efficiency of spectral filtering, and the internal storage efficiency ($\eta_i$). The transmission efficiency includes coupling efficiency between two ends of the waveguide and two optical collimators with single mode fiber-pigtails and a total fiber transmission efficiency in and out of the dilution refrigerator. To measure $\eta_t$, the light centered at 1570 nm - far away from the absorption profile of Er$^{3+}$:LiNbO$_{3}$ - is transmitted into the cryostat at the cooling temperature of 13 mK. Then we evaluated $\eta_t$=26\% by the ratio of output power to input power. The efficiency is 77\% due to spectral filtering of the input photons (5.2 GHz bandwidth) caused by the slightly smaller bandwidth of the AFC memory (4 GHz). Finally, the internal efficiency is calculated as $\eta_i=1.3\times \eta_s/\eta_t$ . For example, when we set the storage time to 200 ns, we detect the coincidence counts triggered by system clocks before (after) storage is 3.4×10$^5$ Hz (2.0×10$^3$ Hz) among input (recalled) signal photons, idler photons and trigger signals for the integrated time of 1000 s. As a result, $\eta_s$ is calculated as 0.59±0.01\% and $\eta_i$ is 2.83±0.06\%. Table S2 shows that the internal storage efficiencies with different storage times range from 100 ns to 240 ns. In the experiment, the storage time is set by changing the properties of the laser for AFC preparation. An internal storage efficiency of 4.74±0.04\% is obtained at a storage time of 160 ns with the magnetic field of 1.3 T, thanks to the $^{93}$Nb caused 12 MHz detuning side holes coincide with the transparency regions of the AFC with a teeth spacing of 6 MHz.
\subsection*{Note S6: The optimization of storage efficiency}
\noindent
The storage efficiency for our memory device is related to the magnetic field and to the pump power for AFC preparation. The magnetic field strength directly influences relaxation dynamics, the lifetime of Zeeman sublevels, side holes detuning, linewidth broadening and so on\cite{ranvcic2018coherence,saglamyurek2015efficient,sinclair2021optical,askarani2019storage}. The power of the AFC pump light greatly affects the non-zero background absorption of AFC by instantaneous spectral diffusion and light-induced two-level systems excitation\cite{thiel2010optical}. To obtain an optimized storage efficiency with the storage time of 200 ns, we optimize the magnetic field and the pump power in our experiment. Figure S5A shows the internal storage efficiency of AFC with the different magnetic field strength. Figure S5B shows the internal storage efficiency of AFC with the different AFC pump power. With an optimized AFC pump power of 28 $\mu$W and an optimized magnetic field of 1.3 T, the highest internal storage efficiency of 2.83±0.06\% is obtained with the storage time of 200 ns.
\subsection*{Note S7: 330 × 330 array of $g_{s,i}^{(2)}(t)$ among 330 temporal modes}
\noindent
For the multimode storage, there are 330 temporal modes generated with a peak separation of 600 ps within a period of 1 $\mu$s. We label each temporal mode of the signal channel as S$\rm_m$ (m=1, 2, ..., 330) and the idler channel as I$\rm_n$ (n=1, 2, ..., 330). By measuring coincidences between S$\rm_m$ and I$\rm_n$, we calculate a 330 × 330 matrix of second-order cross-correlation function. The values of second-order cross-correlation function among 330 single photon modes before/after quantum storage is shown in Fig. S6 (A)/(B). The non-classical nature of the AFC quantum memory and crosstalk between different temporal modes are assessed by the second-order cross-correlation function of correlated modes (m=n) and uncorrelated modes (m$\not=$n), respectively. 

\clearpage
\begin{figure}[ht] 
\centering
\includegraphics[width=16.3cm]{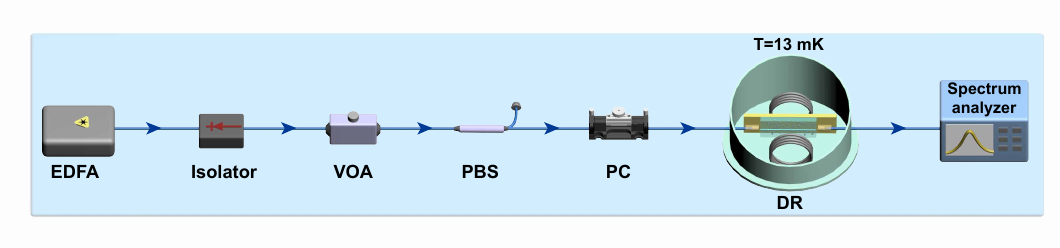}
\caption{\textbf{Experimental setup for measuring the absorption spectrum of Er$^{3+}$:LiNbO$_{3}$.} EDFA: erbium-doped fiber amplifier; VOA: variable optical attenuator; PBS: polarization beam splitter; PC: polarization controller; DR: dilution refrigerator. The Er$^{3+}$:LiNbO$_{3}$ waveguide is placed in a dilution refrigerator with a cooling temperature of 13 mK.}
\label{figS1}
\end{figure}

\clearpage
\begin{figure}[ht]
\centering
\includegraphics[width=16.2cm]{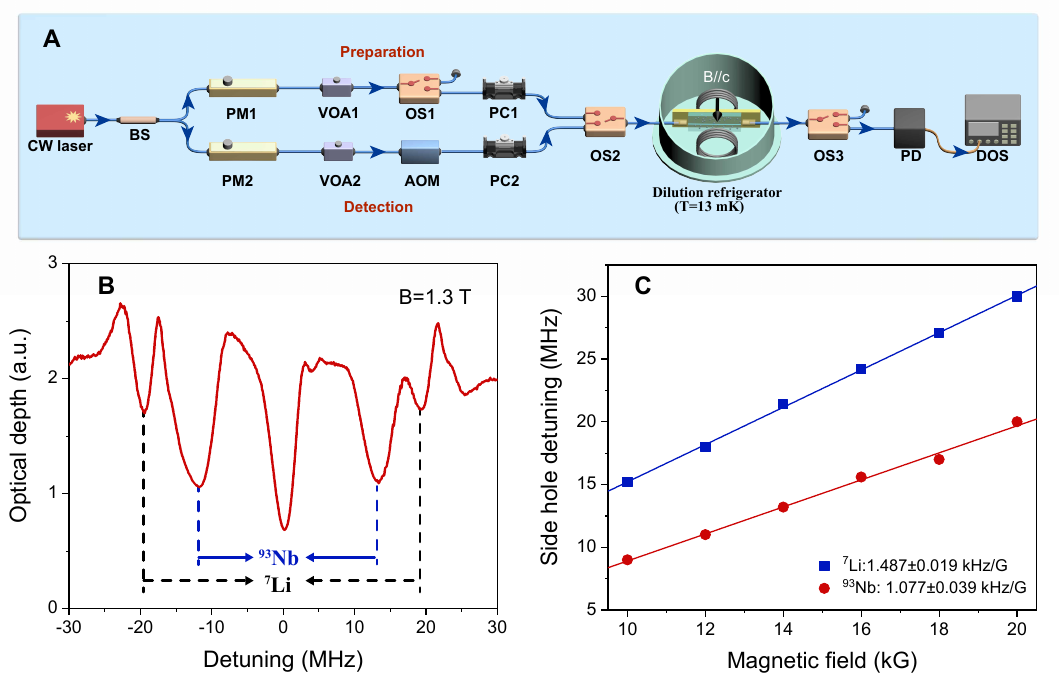}
\caption{\textbf{Spectral hole burning.} (A) The experimental setup. CW laser: continuous-wave laser; BS: beam splitter; PM: phase modulator; VOA: variable optical attenuator; OS: optical switch; AOM: acousto-optic modulator; PC: polarization controller; DR: dilution refrigerator; PD: photodetector; DOS: digital oscilloscope. (B) The optical depth of spectral hole centered at 1532.05 nm in the Er$^{3+}$:LiNbO$_{3}$ waveguide at 13 mK with a magnetic field of 1.3 T along the c-axis. (C) Side hole detunings of $^7$Li and $^{93}$Nb as a function of magnetic field.}
\label{figS2}
\end{figure}

\clearpage
\begin{figure}[ht]
\centering
\includegraphics[width=16.5cm]{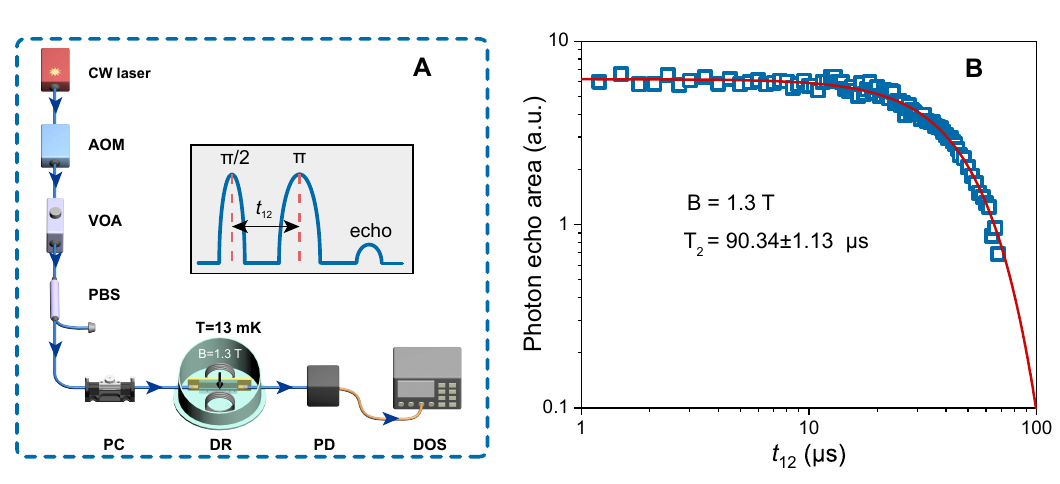}
\caption{\textbf{Measurement of optical coherence time of Er$^{3+}$:LiNbO$_{3}$.} (A) Experimental setup for measuring two-pulse photon echoes. CW laser: continuous-wave laser; AOM: acousto-optic modulator; VOA: variable optical attenuator; PBS: polarization beam splitter; PC: polarization controller; DR: dilution refrigerator; PD: photodetector; DOS: digital oscilloscope. The inset shows the pulses sequence. (B) Results of photon echo area as a function of the time delay $t_{12}$ between the two pulses.}
\label{figS3}
\end{figure}

\clearpage
\begin{figure}[ht]
\centering
\includegraphics[width=16.3cm]{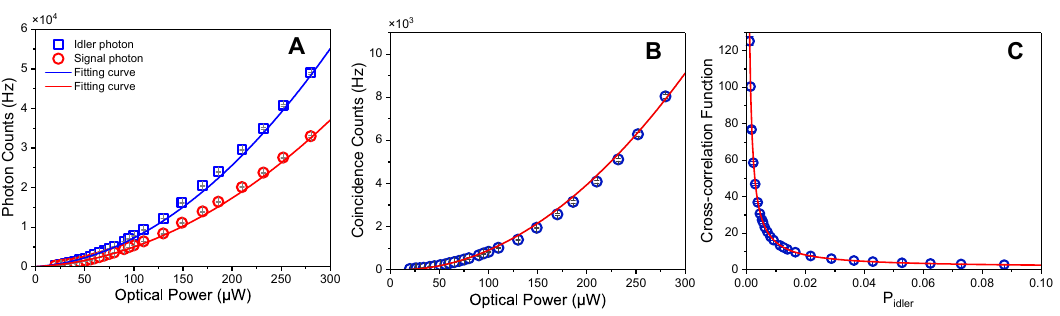}
\caption{\textbf{Results of correlated photon pairs.} (A) Single detection rates of signal photons and idler photons as a function of pump power. The red and blue solid lines are quadratic polynomial fitting curves. (B) Coincidences detection rate between signal photons and idler photons triggered by the system clock as a function of pump power. The red solid line is quadratic polynomial fitting curves. (C) Second-order cross-correlation function as a function of P$\rm_{idler}$. The fitting curve is an inversely proportional function. All error bars are evaluated from the counts assuming Poissonian statistics.}
\label{figS4}
\end{figure}

\clearpage
\begin{figure}[ht]
\centering
\includegraphics[width=16.2cm]{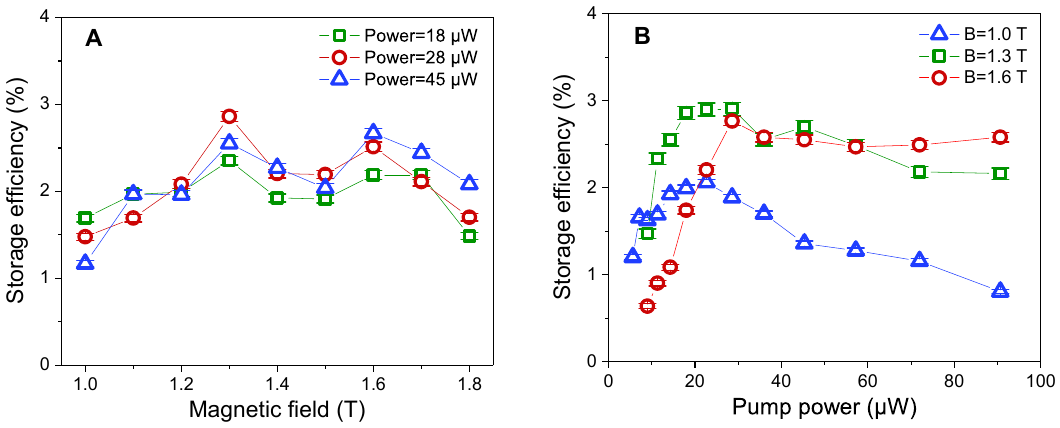}
\caption{\textbf{Calibration measurements to estimate the internal storage efficiency with a storage time of 200 ns.} (A) The internal storage efficiency as a function of the magnetic field. (B) The internal storage efficiency as a function of the pump power. Measurement with a pump time of 200 ms, a storage time of 200 ns and a storage bandwidth of 4 GHz.}
\label{figS5}
\end{figure}

\clearpage
\begin{figure}[ht]
\centering
\includegraphics[width=16cm]{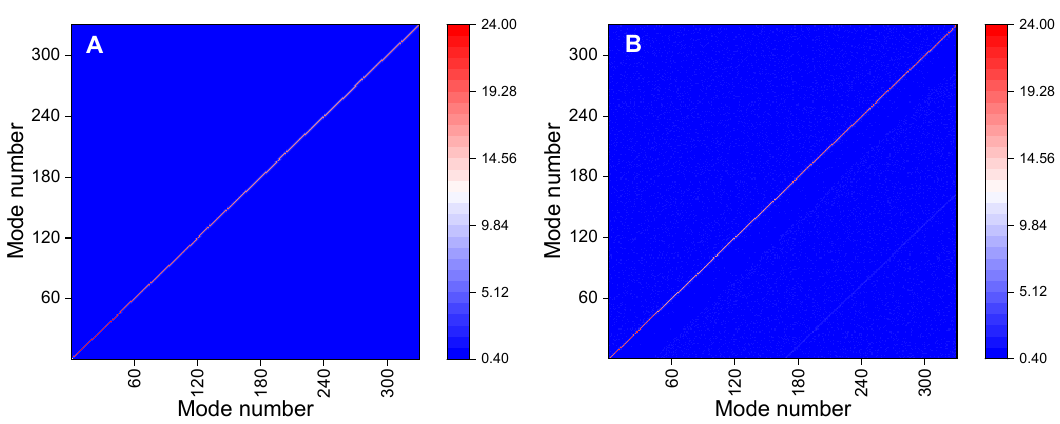}
\caption{\textbf{Calibration measurements to estimate the internal 330×330 array of $g_{s,i}^{(2)}(t)$ among 330 temporal modes.} (A) (B) The values of second-order cross-correlation function among 330 single photon modes before and after quantum storage, respectively.}
\label{figS6}
\end{figure}

\clearpage
\begin{table}\color{black}
\centering
	\caption{Parameters of PPLN module.}
	\label{tables1}
\begin{tabular}{l l c}
\hline
\hline
 Type of waveguide & & RPE waveguide \\
 Length of waveguide & & 50 mm \\
 QPM period & & 19 $\mu$m \\
 SHG normalized conversion efficiency & & 336.07\%/W@1540.56 nm\\
 Length of pigtail & & 20 cm \\
 Input coupling efficiency of PPLN waveguide & & 55.51\% \\
 Output coupling efficiency of PPLN waveguide & & 77.89\% \\
  \hline
  \hline
\end{tabular}
\end{table}

\begin{table}\color{black}
	\caption{Measurement of internal storage efficiency ($\eta$) as a function of the storage time (T$_s$)}
	\label{tables2}
	\begin{ruledtabular}
		\begin{tabular}{c@{}c@{}c@{}c@{}}
			
		T$_S$(ns) & $\eta (\%)$ & T$_S$(ns) & $\eta (\%)$ \\ \hline
			100	& 4.16 ± 0.03 & 180 & 3.39 ± 0.03  \\	
			120	& 2.84 ± 0.03 & 200 & 2.83 ± 0.03  \\
			140	& 3.53 ± 0.03 & 220 & 2.04 ± 0.02  \\
			160	& 4.74 ± 0.04 & 240 & 1.68 ± 0.02 \\
		\end{tabular}
	\end{ruledtabular}
\end{table}

\clearpage

\end{document}